\documentclass[aps,reprint,twocolumn,floatfix,superscriptaddress]{revtex4-1}

\usepackage{mathtools}
\usepackage{amsmath}
\usepackage{amssymb}
\usepackage{bm}
\usepackage{graphicx}
\usepackage{xcolor}
\usepackage{epstopdf}
\usepackage{dcolumn}
\usepackage{bm}

\usepackage{amsfonts,amssymb,amsmath,array}
\usepackage{graphicx}
\usepackage{amsbsy}
\usepackage{color}
\usepackage{hyperref} 
\usepackage{enumerate}
\usepackage{mathtools}
\usepackage{lipsum} 

\newcommand{\lx} {\left}
\newcommand{\rx} {\right}

\newcommand{\ave}[1] {\lx\langle #1 \rx\rangle}

\newcommand{\om} {\omega}

\begin{document}

\title{Revealing the non-equilibrium nature of a granular intruder:\\ the crucial role of non-Gaussian behavior}
\author{D. Lucente}
\affiliation{Department of Physics, University of Rome Sapienza, P.le Aldo Moro 2, 00185, Rome, Italy}
\affiliation{Institute for Complex Systems - CNR, P.le Aldo Moro 2, 00185, Rome, Italy}

\author{M. Viale}
\affiliation{Department of Physics, University of Rome Sapienza, P.le Aldo Moro 2, 00185, Rome, Italy}
\affiliation{Institute for Complex Systems - CNR, P.le Aldo Moro 2, 00185, Rome, Italy}

\author{A. Gnoli}
\affiliation{Department of Physics, University of Rome Sapienza, P.le Aldo Moro 2, 00185, Rome, Italy}
\affiliation{Institute for Complex Systems - CNR, P.le Aldo Moro 2, 00185, Rome, Italy}

\author{A. Puglisi}
\affiliation{Department of Physics, University of Rome Sapienza, P.le Aldo Moro 2, 00185, Rome, Italy}\affiliation{Institute for Complex Systems - CNR, P.le Aldo Moro 2, 00185, Rome, Italy}
\affiliation{INFN, University of Rome Tor Vergata, Via della Ricerca Scientifica 1, 00133, Rome, Italy}

\author{A. Vulpiani}
\affiliation{Department of Physics, University of Rome Sapienza, P.le Aldo Moro 2, 00185, Rome, Italy}

\begin{abstract}
The characterization of the distance from equilibrium is a debated problem in particular in the treatment of experimental signals. If the signal is a 1-dimensional time-series, such a goal  becomes challenging. A paradigmatic example is the angular diffusion of a rotator immersed in a vibro-fluidized granular gas. Here, we experimentally observe that the rotator's angular velocity exhibits significative differences with respect to an equilibrium process. Exploiting the presence of two relevant time-scales and non-Gaussian velocity increments, we quantify the breakdown of time-reversal asymmetry, which would vanish in the case of a 1d Gaussian process. We deduce a new model for the massive probe, with two linearly coupled variables, incorporating both Gaussian and Poissonian noise, the latter motivated by the rarefied collisions with the granular bath particles. Our model reproduces the experiment in a range of densities, from dilute to moderately dense, with a meaningful dependence of the parameters on the density. \textcolor{black}{We believe the framework proposed here opens the way to a more consistent and meaningful treatment of out-of-equilibrium and dissipative systems.}
\end{abstract}
\maketitle
{\em Introduction.} Non equilibrium systems, even after the seminal contribution of the
pioneers in the last century, still represent a challenging frontier
of statistical mechanics~\cite{KTH91}. For sure the archetypal example of non
equilibrium is  Brownian Motion and its formalization in terms of
equations of motion with a random force~\cite{db05}. Following this original starting point, many
phenomena have been modelled in terms of stochastic differential
equations, in particular continuous stochastic processes involving
white noise~\cite{R89}. In addition, more general Markov processes (e.g. Master
equations) have been used, particularly for biological and chemical
systems~\cite{G90}.\\
In the last decades stochastic thermodynamics entered the scene, a new approach in terms of Markov
processes which attempts to formalise concepts such as work, heat and entropy for mesoscopic
systems~\cite{BPRV08,Sekimoto2010,seifertrev,pelipigo}. While its theoretical framework can be considered basically mature,
the treatment of data coming from experiments remains a debated problem.
For instance it is not always obvious how to infer, from experimental signals,
relevant features such as the equilibrium or non-equilibrium
nature of the system~\cite{seifert2019stochastic,skinner21,van2022thermodynamic,lucente2022inference,harunari2022learn}.  In addition, in several non-trivial
situations, it is not straightforward how to follow Langevin's
path in order to achieve an appropriate mathematical modelling of
the system under investigation~\cite{baldovin2018role,baldovin2019langevin}.
Among non-equilibrium systems, granular gases demonstrated to be
particularly interesting~\cite{PL01}. In fact, they are experimentally accessible and, being non-Hamiltonian and
dissipative, constitute an intriguing  theoretical challenge~\cite{puglio15}.\\
Here, we present an analysis of experimental data obtained from a
vibrofluidized granular setup~\cite{scalliet2015cages}. A massive probe is suspended in the granular
gas and, under the effect of the collisions, it performs a rotational motion. The aim of this Letter is to investigate the statistical features of the probe's motion  \textcolor{black}{for inferring the properties of the granular system}, in
particular to shed light on the non-equilibrium nature of the system
as well as its modelling in terms of a suitable stochastic
process. \textcolor{black}{We stress that the non-equilibrium nature of the system is indisputable (for instance it lacks equipartition of energy~\cite{feitosa2002breakdown}) . Nonetheless}, previous attempts have shown that, both at the experimental and theoretical level, revealing this nature is particularly difficult when looking only at the isolated tracer dynamics, i.e. without resorting to the study of correlations with the surrounding medium~\cite{sarracino2010irreversible} or by perturbing the experiment with an external force~\cite{gnoli2014nonequilibrium}.
\begin{figure}[ht!]
    \includegraphics[width=0.48\textwidth]{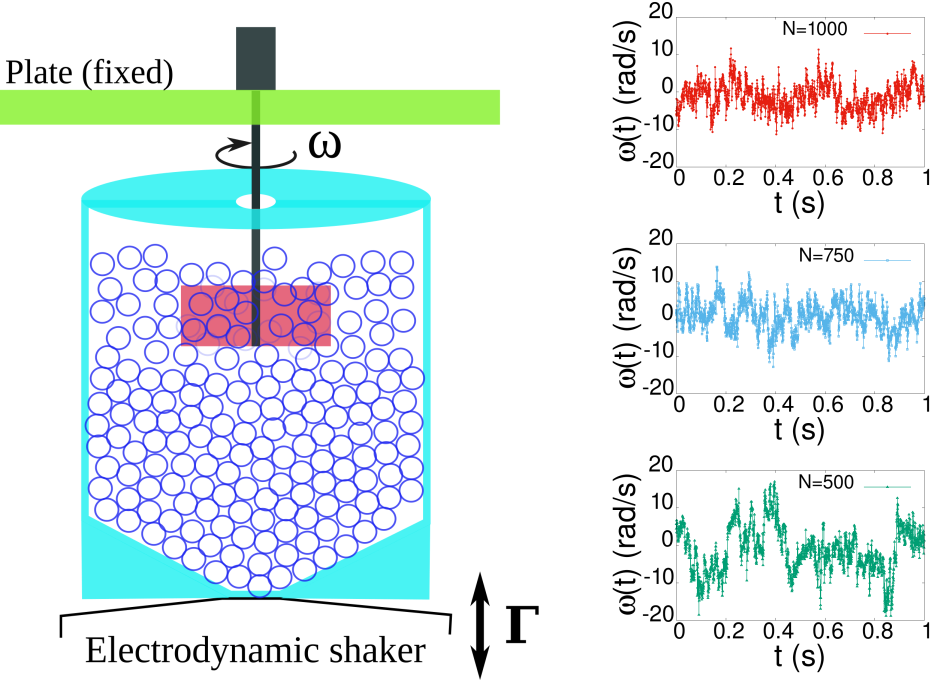}
  \caption{Scheme of the experimental setup (left) and time series of the probe's angular velocity $\omega(t)$ (right) for different values of the number of spheres $N$.
  }
  \label{fig:setup}
\end{figure}
Our analysis shows that the system
exhibits both non-Gaussian and non-equilibrium properties. Therefore, a
description in terms of linear differential equation with Gaussian white noise
lacks important features of the underlying dynamics. This can appear 
surprising, in view of several studies of driven granular gases, where
the usual linear Langevin equation has been found successful for the
description of other features of massive probes, particularly its several diffusion regimes~\cite{lasanta2015itinerant,baldovin2019langevin,plati2019dynamical,plati2020slow}. An important
outcome of our study is that even remaining in the context of linear
models, the introduction of a suitable non-Gaussian noise is
sufficient to catch the non equilibrium statistical properties of the system. \textcolor{black}{We expect such an ingredient to be relevant not only for granular systems but for other kinds of macroscopic 'baths', e.g. in active matter~\cite{maggi2014generalized}.}\\
{\em Setup.}
The setup used here is an improved version of the one used in~\cite{scalliet2015cages}, see also~\cite{d2003observing,naert2012experimental}. The
granular medium made of $N$ spheres ($N=\{500,750,1000\}$) of diameter $d=4$mm is placed
in a cylindrical container of volume $\sim 7300$ times that of a
sphere (the average \textcolor{black}{volume fraction} is therefore in the range $\sim 7-14 \% $). The
container is vertically shaken with a sinusoidal signal whose amplitude and frequency are $A=1.6$mm and $f_{ext}=53$ Hz. 
A blade, our ``massive tracer'' with cross section $\sim 32\text{mm} \times 5\text{mm}$, is suspended into the granular medium and rotates around a vertical axis  \textcolor{black}{$z$ (see left panel of Fig.~\ref{fig:setup})}. Its angular velocity $\omega(t)$ and the traveled angle of rotation $\theta(t)=\int_0^t \omega(t')dt'$ are measured with a time-resolution of $2$ kHz. The blade, interacting with the spheres, performs a motion qualitatively similar to an angular Brownian motion \textcolor{black}{(right panels Fig.~\ref{fig:setup})}. The shaking intensity is measured by the normalized mean squared acceleration of the vibrating plate $\Gamma=\sqrt{2\langle \ddot{z}^2 \rangle}/g \simeq 18$ \textcolor{black}{where $g$ is the acceleration of gravity}.
\begin{figure}[ht!]
\includegraphics[width=0.48\textwidth]{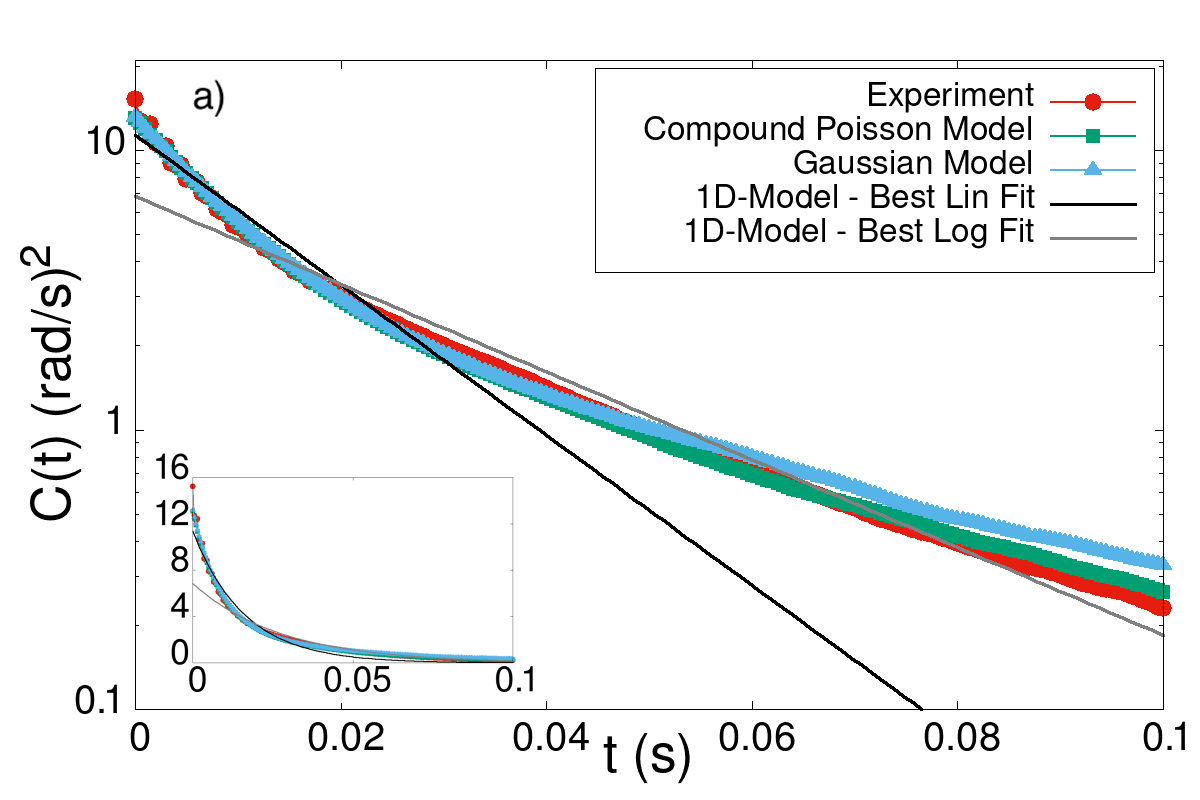}
\includegraphics[width=0.48\textwidth]{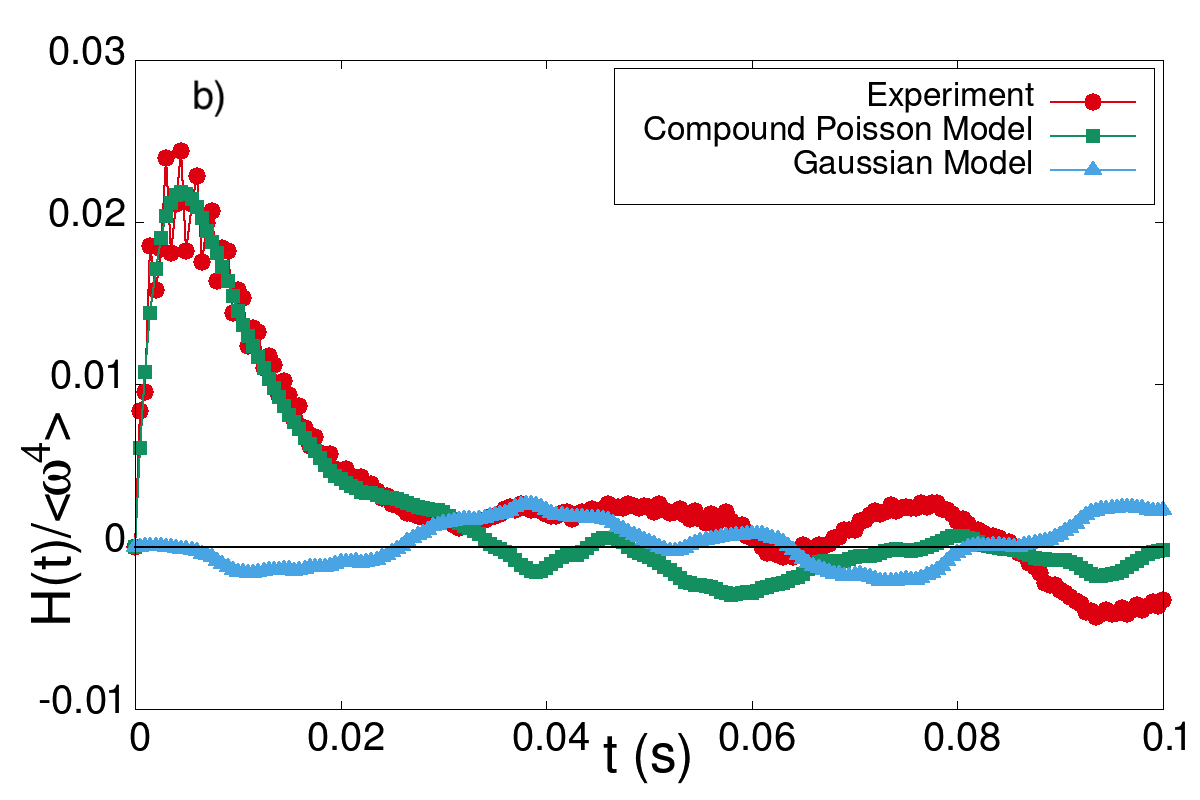}
\caption{
Comparison between experiment (red) and models (blue and green) in the case of $N=1000$: time correlation function $C(t)$ (\textcolor{black}{a}) and $H(t)$ (\textcolor{black}{b}). Numerical integration of Eq.~\eqref{eq:model} is done with two different choices for the noise $\xi_1$: it is a Poisson compound noise (green line) or a white noise (blue line). The model parameters were obtained by fitting in linear scale the experimental curves for $C(t)$ and $H(t)$ (or only $C(t)$ in the case of the Gaussian simulation) with the expressions in \eqref{eq:params}. By looking at the best fit with $C(t)=Ae^{-at}$ both in linear (inset) and logarithmic scale, one notes the impossibility to mimic the experimental data with a single time scale. All simulation data are been obtained with the same statistics of experimental data in order to appreciate the relevance of the finite-time fluctuations.}
\label{fig:modsim}
\end{figure}
\begin{figure}[ht!]
\includegraphics[width=0.48\textwidth]{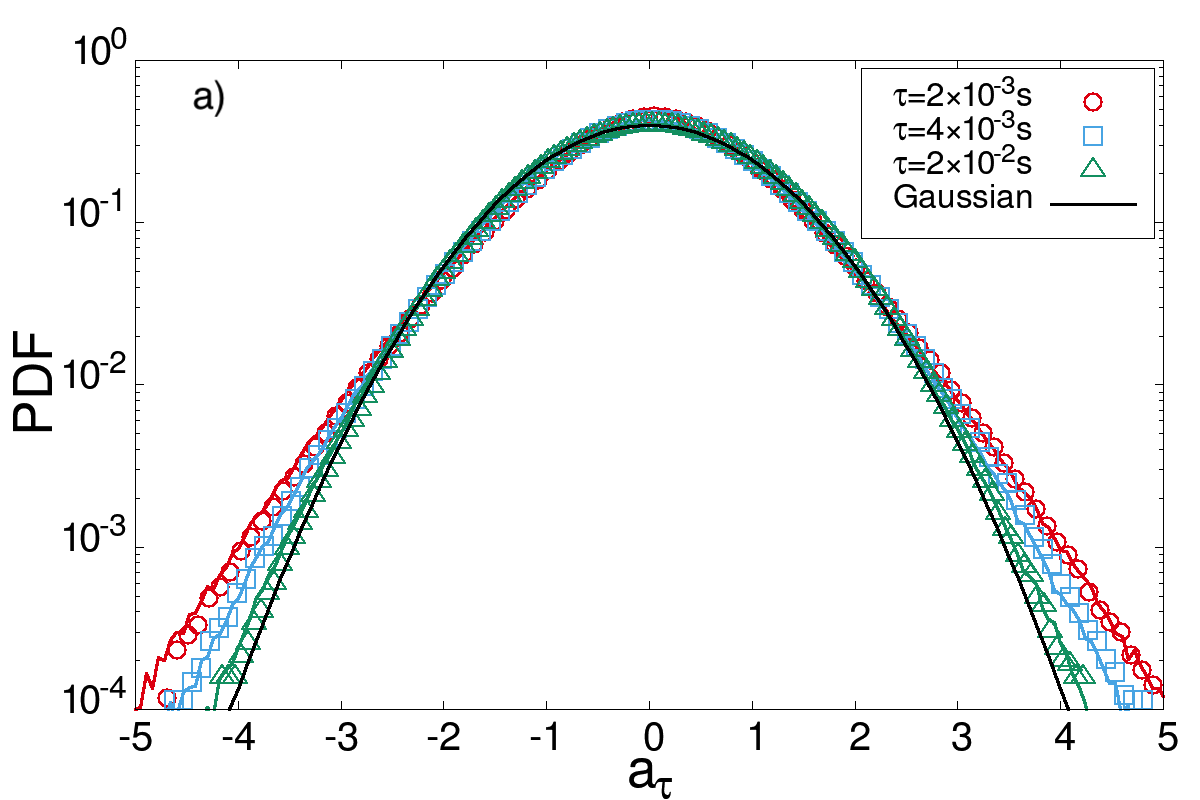}\\
\includegraphics[width=0.48\textwidth]{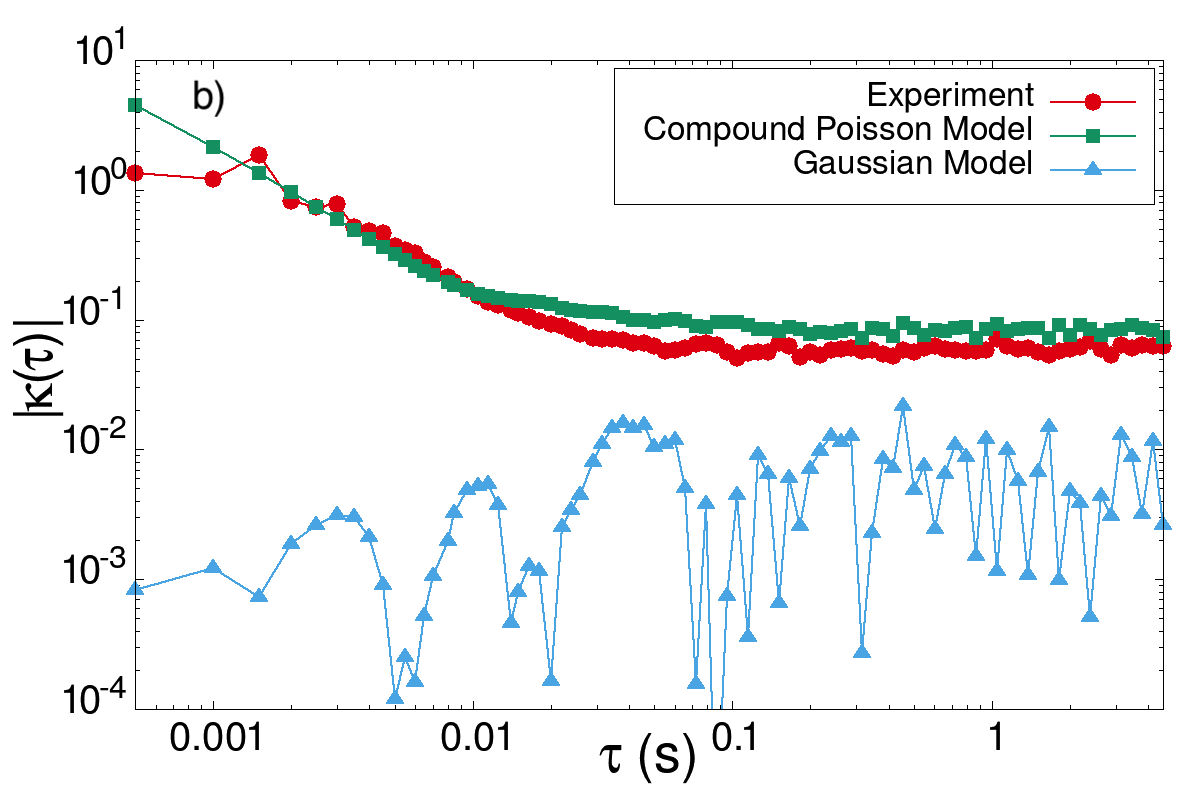}
 \caption{Statistical distribution of the standardized angular velocity increments \textcolor{black}{$a_\tau(t)=\Delta_\tau\omega /\sqrt{\ave{\lx(\Delta_\tau\omega\rx)^2}}$}: comparison between models and the experiment with $N=1000$. In \textcolor{black}{panel a)} we show the distributions of the variables at three different lag-times $\tau$: the empty symbols are experimental data, while solid curves are the interpolation of the numerical integration of the model in Eq.~\eqref{eq:model} with Poisson compound noise $\xi_1$. The black line is the normal distribution for reference.  \textcolor{black}{Panel b)} shows the absolute value of the excess kurtosis $\kappa(\tau)$ as a function of $\tau$ for experimental measurements (red circles) and numerical simulations with $\xi_1$ chosen to be a compound Poisson noise (green squares) or white noise (blue triangles).}
 \label{fig:pdf}
\end{figure}\\
{\em Results.}
The time correlation function of the angular velocity $C(t)=\ave{\om(t)\om(0)}$ contains information on the relevant time scales of the system, \textcolor{black}{where $\ave{\cdot}$ denotes ensemble averages and can be replaced by temporal averages over long trajectories assuming stationarity and ergodicity}. 
At first glance, it might seem that the correlations decay exponentially with a single relaxation time. Since $C(t)\to 0$ fast enough when $t\to \infty$, i.e. $\int_0^{\infty} C(t) dt < \infty$, at large times the angle $\theta$ of the blade performs a standard diffusion, i.e. $\ave{|\theta(t)-\theta(0)|^2}\sim t$. One might be tempted to describe the system with a simple one-dimensional Gaussian model, i.e. $\dot{\omega} + \eta \omega = \xi$ \textcolor{black}{where $\eta$ is the damping and $\xi$ is a Gaussian white noise}. 
However, a closer inspection reveals that
the system has at least two relevant time scales, see Fig.~\ref{fig:modsim} \textcolor{black}{and its inset}. Thus such a model with a single time scale is inadequate to reproduce the $C(t)$. This is not enough, since "Fickian yet non Gaussian diffusion" has been observed in many physical system ranging from colloidal systems to super-cooled liquids~\cite{wang2012brownian,kanazawa2020loopy,pastore2022model,rusciano2022fickian}, therefore it is crucial to investigate the signal statistics by calculating higher order moments.
Moreover, the non-Gaussian statistics of the signal allows one to deduce information about the time reversal symmetry of the system.
In fact, as discussed in~\cite{lucente2022inference}, a one-dimensional Gaussian signal is always invariant under time reversal, even if produced by an out-of-equilibrium physical system. 
On the contrary, if its statistics is non-Gaussian, the signal may break time reversal symmetry, and therefore one may quantify the distance from equilibrium of the underlying physical system. 
Often, the entropy production is used for quantifying temporal asymmetries, but it is a notoriously difficult quantity to measure, given the large amount of data required to obtain accurate estimates~\cite{li2019quantifying}. Therefore, we take an alternative route by using higher order correlation function for detecting temporal asymmetries.
Indeed, a system is at equilibrium if $C_{fg}(t)=\ave{f(t)g(0)}=C_{gf}(t)=C_{fg}(-t)$ for all functions $f$ and $g$. 
Thus, as proposed by Pomeau~\cite{pomeau1982symetrie}, if observables $f$ and $g$ exist, such that $C_{fg}(t) \neq C_{fg}(-t)$, the difference $C_{fg}(t)-C_{fg}(-t)$ can quantify the distance from equilibrium. We find that the simplest functions that provide a non-trivial results are $f=\omega$ and $g=\omega^3$.
As shown in Fig.~\ref{fig:modsim} \textcolor{black}{b)}, $H(t)=C_{\om\om^3}(t)-C_{\om\om^3}(-t)$ is significantly different from zero for short times $t$, indicating that the underlying dynamics is out of equilibrium. 
Further corroboration of the non-Gaussianity of the system is obtained by looking at the statistics of the \textcolor{black}{standardized} angular velocity increments 
\textcolor{black}{$a_\tau(t)=\Delta_\tau\omega /\sqrt{\ave{(\Delta_\tau\omega)^2}}$ ($\Delta_\tau\omega=\lx[\om(t+\tau)-\om(t)\rx]$)} by varying the lag-time $\tau$.
Note that, as in turbulence, derivatives or other filters are widely used to highlight the statistical properties of a system~\cite{frisch1981intermittency}.
It has been found \cite{baldovin2019langevin} that the probability density function (pdf) of the probe's velocity is Gaussian-like but the pdfs of the increments $a_\tau$ are not for small values of $\tau$ and become Gaussian-like for \textcolor{black}{$\tau > 10^{-2} s$}, as shown in Fig.~\ref{fig:pdf} \textcolor{black}{a)} In particular, by looking at the excess kurtosis \textcolor{black}{$\kappa(t)=\lx(\ave{a_\tau^4}/\ave{a_\tau^2}^2-3\rx)$ we note that it is much larger than the statistical fluctuations due to the finiteness of the observations as shown in Fig.~\ref{fig:pdf} \textcolor{black}{b)} and that is close to zero, corresponding to the Gaussian-like case, just at $\tau > 10^{-2}s$.}
An effective model should therefore be able to account for: multiple time scales, standard diffusion at large times, Gaussian-like velocity pdf, non-Gaussianity of velocity increments and time reversal asymmetry.

{\em Model.} 
Typically, the information available to an observer does not allow to adopt a prescribed protocol to determine a model. Therefore it is necessary to assume that a certain description is relevant (at a given scale) and check a posteriori its validity. 
The presence of multiple time scales as well as non Gaussianity could be explained by considering non-linear 1d models driven by white noise processes. However, such 1d models can not be out-of-equilibrium, unless with periodic boundary conditions, and therefore can not reproduce one of the main observed property, that is the aforementioned lack of time reversal symmetry of the angular velocity signal. Thus, a meaningful effective model should have at least two degrees of freedom~\cite{lucente2022inference}. 
Among the possible 2d systems, we choose to focus our attention on linear systems for two reasons. The first is that they have already been shown to correctly reproduce many properties of this system \cite{scalliet2015cages,lasanta2015itinerant,baldovin2019langevin,plati2020slow}. The second one is related to the observation that $\ave{(\theta-\theta_0)^2}\sim t$ even if $\omega$ is not Gaussian.  Popular theoretical frameworks to explain this phenomenon rely on superstatistics such as diffusing diffusivity or, differently, on Continuous Time Random Walk (CTWR)\cite{chubynsky2014diffusing,chechkin2017brownian,barkai2020packets}. But we note that every linear system driven by a delta-correlated noise shows a standard diffusion, regardless of the distribution density of the stochastic forcing. There is of course a third clue about the linearity of the process, which is the observation of a correlation function which decays exponentially in time. None of these arguments is compelling, but together they are a strong hint in  favour of the choice of a linear process. We also note that in \cite{baldovin2019langevin} the linearity of the model is not assumed {\it a priori} but obtained from the analysis of the signal. Of course in an Occam Razor based approach one should prefer a linear model (which has less parameters than non-linear ones) if there are no evidences for non-linearities. 
In the following, we propose a linear non-Gaussian stochastic process and give evidence that it is suitable to model the vibrofluidized granular gas experiment investigated in this Letter. 
Let us consider the following linear stochastic differential equation
\begin{equation}
\label{eq:model}
\lx\{
\begin{aligned}
\dot{\omega}+ \gamma \omega &= \Omega + \xi_1 \\
\dot{\Omega}+ \mu \Omega &= \xi_2 \\
\end{aligned}
\rx.
\end{equation}
\textcolor{black}{where $\gamma$ is a damping coefficient for the probe and arises both from
the average effect of collisions with the granular gas as well as with other sources of
viscosity (air, solid friction, etc.), the variable $\Omega$ is a collective
variable which takes into account the inertial effect of the surrounding granular medium and $1/\mu$ is its typical relaxation timescale.}
The fluctuating force $\xi_2(t)$ is a typical white noise \textcolor{black}{representing fluctuations of $\Omega$ ($\ave{\xi_2(t)}=0$ and $\ave{\xi_2(t)\xi_2(t')}=\sigma_2^2\delta(t-t')$), while for $\xi_1(t)$ we will consider two cases: a white Gaussian noise or a compound Poisson process, i.e. $\xi_1(t)=\sum_j z_j\delta(t-t_j)$ (where $\delta(t)$ is a Dirac delta) with intensity $\lambda$ and normally distributed jumps}. This means that the time intervals between jumps $\Delta_j=t_j-t_{j-1}$ are independent of each other and exponential distributed ($P(\Delta) \sim \lambda e^{-\lambda \Delta}$), while the amplitude of the jumps $z_j$ is sampled from a normal distribution ($P(z) \sim \mathcal{N}_\sigma(z)$) with zero mean and standard deviation $\sigma$.
Note that $\ave{\xi_1^4}_c=\textcolor{black}{\ave{\xi_1^4}-3\ave{\xi_1^2}^2=}3\lambda \sigma^4$ and that $\xi_1$ tends to a standard Gaussian white noise \textcolor{black}{($\ave{\xi_1(t)}=0$ and $\ave{\xi_1(t)\xi_1(t')}=\sigma_1^2\delta(t-t')$)} in the limit $\lambda \to \infty$, $\sigma \to 0$ with $\lambda \sigma^2 =\sigma_1^2 = constant$. The use of such noise has both a physical and a mathematical justification. From a physical point of view, $\xi_1$ can be interpreted as the process originating from instantaneous collisions of granular particles with the blade. A rigorous justification for the model in Eq.~\ref{eq:model} could be obtained by the design of a kinetic theory for inelastic hard spheres including the specific setup of our experiment. Recently, it has been shown that non-Gaussian white noises like $\xi_1$ can be derived from microscopic theories through a systematic expansion of the Boltzmann-Lorentz equation governing the evolution of the blade~\cite{kanazawa2015minimal,kanazawa2015asymptotic,kanazawa2017statistical}. In our opinion those studies offer a general justification for non-Gaussian white noise in vibrated and diluted granular experiments. However, this reasoning cannot explain the existence of the second degree of freedom $\Omega$. We conjecture that the missing ingredient in those previous theory is the interaction between the granular bath and the particular boundary conditions in our setup, which prevent the rapid thermalization of the tangential components of the velocities of the spheres, which is likely to be responsible for the memory here modelled in terms of $\Omega$. 
In addition, even from a mathematical point of view, by virtue of the Levy-Ito decomposition theorem, the used noise has an interesting structure since it is one of the three contributions to process with independent and identically distributed increments~\cite{ken1999levy,kyprianou2014fluctuations,schilling2016introduction}.
As shown in \cite{kanazawa2015minimal,kanazawa2015asymptotic,kanazawa2017statistical, kusmierz2018thermodynamics,bialas2020colossal}, a system driven by Levy noises (like $\xi_1$) is necessarily out of equilibrium. Thus, it should be sufficiently general to capture, at least qualitatively, the relevant features of the granular gas. 
For our model $C(t)$ and $H(t)$ are given by
\begin{eqnarray}
C(t) = A e^{-\gamma t} + B e^{-\mu t} ,\quad
H(t) = D\lx(e^{-\gamma t}-e^{-3\gamma t}\rx) \nonumber\\
     A = \lambda \sigma^2 - B, \quad 
     B = \frac{\sigma_2^2}{\gamma^2-\mu^2} ,\quad
     D = \frac{3\lambda \sigma^4}{4\gamma} \label{eq:params}
\end{eqnarray}
\begin{figure}[t!]
\includegraphics[width=0.48\textwidth]{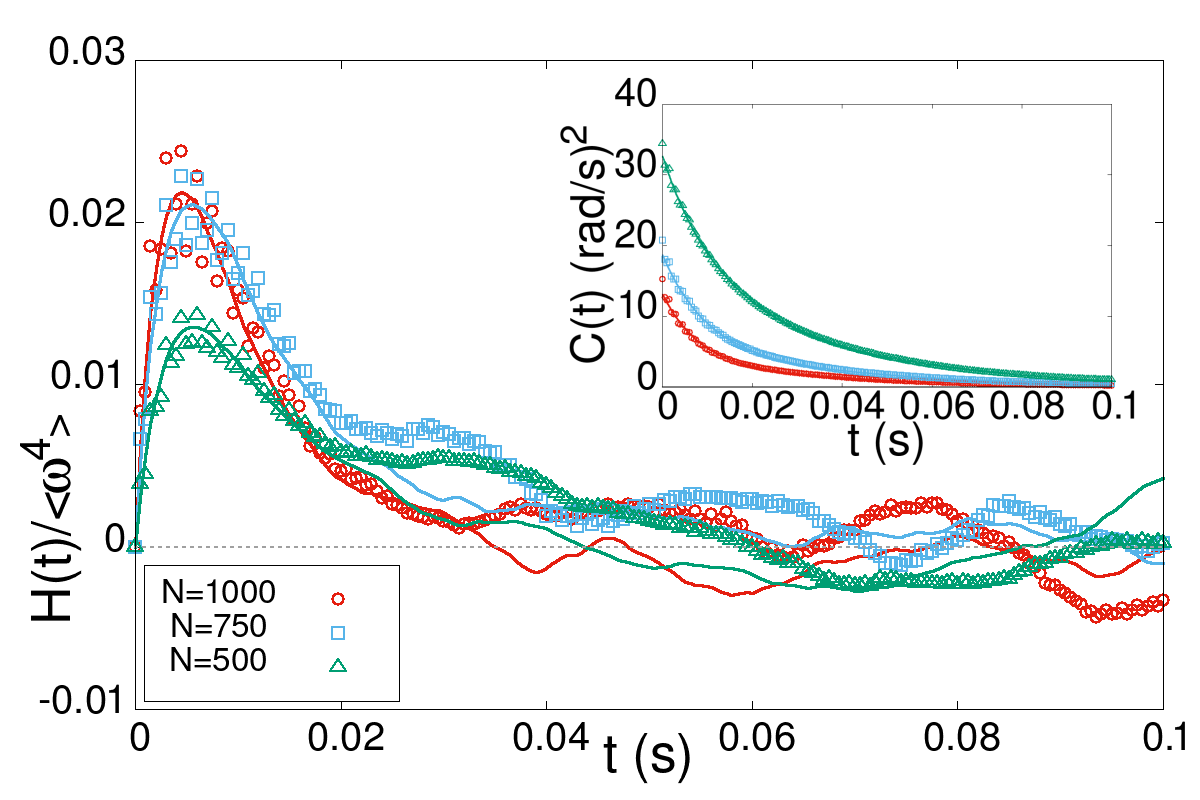}
 \caption{Comparison between experiments with different $N$.
 The empty symbols are experimental data, while solid curves come from numerical integration of the model in Eq.~\eqref{eq:model} with Poisson process for $\xi_1(t)$. The inset shows the corresponding correlation functions.}
 \label{fig:exp}
\end{figure}
\begin{table}[h]
\begin{tabular}{c|c|ccccc}
$h (mm)$ & $N$ & $\gamma\,\lx(s^{-1}\rx)$	& $\mu\,\lx(s^{-1}\rx)$ & $\sigma_2 \lx(s^{-5/2}\rx)$ & $\sigma (s^{-1})$	& $\lambda\,(s^{-1})$ \\
\hline
5  & 1000 & 122.9	& 26.17 & 1647	& 1.415 & 1261 \\
5  & 750  & 105.3	& 25.13 & 1776	& 1.631 & 1116 \\
5  & 500  & 96.04	& 29.35 & 2965	& 1.943 & 1024 \\
\hline
15 & 500  & 107.5	& 22.41 & 1880	& 1.058 & 1617 \\
\end{tabular}
\caption{Table with the best fit parameters (the errors are order $5\%-10\%$) for the different experiments. The values are those used for numerical simulations. }
\label{tab:fitres}
\end{table}
These expression can be employed to infer the model parameters from the experimental results. For instance, to get a good and robust match of both curves without fitting too many parameters at once, we infer the inverse relaxation time $\gamma$ and the product $3\lambda \sigma^4=4 D \gamma$ from $H(t)$, then, once $\gamma$ fixed, we can fit $C(t)$ via $A,B$ and $\mu$ so, by inverting the expressions in \eqref{eq:params}, we get the remaining parameters.\\
In Fig.~\ref{fig:modsim} we verify that the numerical integration of the model is able to reproduce experimental results when $\xi_1(t)$ is a compound Poisson noise, while the choice of $\xi_1(t)$ to be white noise  only  reproduces the time correlation of the system, while it is inadequate to predict the non-equilibrium nature of the real process.\\
We can check the consistency of the model by comparing the distribution of the increments $a_\tau$ and the behaviour of $\kappa(\tau)$ as a function of $\tau$. Fig.~\ref{fig:pdf} \textcolor{black}{a)} shows a very good agreement in a wide range of lag-times: the model correctly reproduces the pdfs as well as their behavior for \textcolor{black}{$\tau\gg 1/\lambda \sim 10^{-3} s$}.
Regarding $\kappa(\tau)$, it can be seen from Fig.\ref{fig:pdf} \textcolor{black}{b)} that the model has a little discrepancy in the long times behavior but matches well enough the functional shape of experimental curve in almost all range of lag-times \textcolor{black}{and moreover correctly predicts its relaxation timescale $\tau\sim 1/\gamma \sim 10^{-2} s$.}\\
To conclude, we note that the model reproduces quite faithfully the dynamics of the system as the experimental conditions vary, as can be seen from Fig.~\ref{fig:exp} showing $C(t)$ and $H(t)$ for different number of particles $N$. The upper rows of Table~\ref{tab:fitres} reports the parameters of the model obtained with the procedure explained above \textcolor{black}{by varying $N$}. We observe that \textcolor{black}{the observed trends of the parameters follow a simple physical interpretation in agreement with our expectations.
Indeed, since in all experiment the shaker injects a similar amount of energy (weakly dependent upon the number of particles) which is then dissipated through the collisions, increasing $N$ leads to a lower kinetic energy per particle and therefore lower values for the noise amplitudes $\sigma$ and $\sigma_2$. On the contrary, since the number of collision increases both the damping coefficient $\gamma$ and the collision rate $\lambda$ are observed to increase.}
\textcolor{black}{Moreover, in order to check whether our physical interpretation is consistent, we performed additional experiments with a different blade (h = 15mm) in the most diluted case ($N=500$). Having increased the cross section of the probe we expect a greater number of collisions between particles and blade which should lead to an increase of the damping coefficient $\gamma$ and of the collision rate $\lambda$ on the one hand, and in a decrease of the noise amplitudes $\sigma$ and $\sigma_2$ on the other. These expectations are confirmed by the experimental results, as can be seen by comparing the last two rows of Table~\ref{tab:fitres}.}

{\em Conclusions.}
We have analysed out-of-equilibrium Brownian-like motion in an experiment with a rotating tracer immersed in a vibro-fluidized granular medium. We provide a detailed quantification of how far from equilibrium such a system is by just looking at the $\om$ signal of the tracer. Until now such a task was only possible through the use of auxiliary observables or specific experiments such as perturbation-response experiments. Careful examination of the angular velocity time-series revealed clear non-equilibrium features in the shape of non-Gaussian velocity increments and asymmetric time-correlations. These observations led us to propose a model for the tracer's dynamics incorporating Poissonian (non-Gaussian) noise, coherent with the physical intuition of a dynamics where collisions are sparse in time. We underline that for this model entropy production diverges, since most of the trajectories have no time-reversed counterpart~\cite{CP15}. \textcolor{black}{Finally, we stress that the techniques we used
to analyse experimental signals are absolutely general and  can be applied for modeling other systems well beyond the realm of granular
material. In addition, these techniques open the perspective of extending the general procedure used in \cite{baldovin2019langevin} to derive Langevin equations also for non-Gaussian noises.} 
\begin{acknowledgments}
A.P., A.G., D.L. and A.V. acknowledge the financial support from the MIUR PRIN 2017 (project "CO-NEST" no. 201798CZLJ). M.V. and D.L. were supported by ERC-2017-AdG (project "RG.BIO" no. 785932). M.V. was supported by MIUR FARE 2020 (project "INFO.BIO" no. R18JNYYMEY).
\end{acknowledgments}
\bibliographystyle{apsrev4-1}
\bibliography{biblio}
\end{document}